\documentclass[11pt,a4paper]{article}

\usepackage{graphicx}
\usepackage[myheadings]{fullpage}
\usepackage{epsfig,epsf}
\usepackage{amssymb}
\usepackage{setspace}
\usepackage{amsmath}

\usepackage{txfonts}
\usepackage{apacite}
\usepackage{xcolor}
\usepackage{amsfonts}
\usepackage{dsfont}

\newcommand{\change}[1]{#1}
		
\begin{document}

\title{Joint modelling of repeated multivariate cognitive measures and competing risks of dementia and death:
 a latent process and latent class approach}

\author{C\'ecile Proust-Lima $^\text{*,a,b}$, Jean-Fran\c cois Dartigues $^\text{a,b}$ and H\'el\`ene Jacqmin-Gadda $^\text{a,b}$}

\date{May 2015}

\maketitle

\vspace{0.3cm}

\noindent 
$^\text{a}$ INSERM, U897, Epidemiology and Biostatistics Research Center, 
F-33076 Bordeaux, France \\
$^\text{b}$ Universit\'e de Bordeaux, ISPED, 
F-33076 Bordeaux, France \\
$~~$\\
$^\text{*}$ correspondence to: \\
$~~~$ C\'ecile Proust-Lima, Inserm U897, ISPED, 146 rue Léo Saignat, 33076 Bordeaux Cedex, France. \\
$~~~$ E-mail: cecile.proust-lima@inserm.fr

\vspace{0.9cm}
\noindent {\bf {Abstract:}} Joint models initially dedicated to a single longitudinal marker and a single time-to-event need to be extended to account for the rich longitudinal data of cohort studies. Multiple causes of clinical progression are indeed usually observed, and multiple longitudinal markers are collected when the true latent trait of interest is hard to capture (e.g. quality of life, functional dependency, cognitive level). These multivariate and longitudinal data also usually have nonstandard distributions (discrete, asymmetric, bounded,...). We propose a joint model based on a latent process and latent classes to analyze simultaneously such multiple longitudinal markers of different natures, and multiple causes of progression. A latent process model describes the latent trait of interest and links it to the observed longitudinal outcomes using flexible measurement models adapted to different types of data, and a latent class structure links the longitudinal and the cause-specific survival models. The joint model is estimated in the maximum likelihood framework. A score test is developed to evaluate the assumption of conditional independence of the longitudinal markers and each cause of progression given the latent classes. In addition, individual dynamic cumulative incidences of each cause of progression based on the repeated marker data are derived. The methodology is validated in a simulation study and applied on real data about cognitive aging obtained from a large population-based study. The aim is to predict the risk of dementia by accounting for the competing death according to the profiles of semantic memory measured by two asymmetric psychometric tests.\\


\noindent {\bf {Funding:}} Agence Nationale de la Recherche [grant 2010 PRSP 006 01]\\

\pagebreak

\doublespace

\section{Introduction}
\label{sec1}
With the large development of cohort studies and the collection of repeated markers and clinical events, joint models for repeated measurements and time-to-events are now widespread in the biostatistics community \cite{
tsiatis2004,rizopoulos2012,proust-lima2014}. By focusing on the joint distribution of the different types of outcomes, these models provide a general framework to better describe the link between continuous progression of diseases through longitudinal outcomes such as biomarkers or more generally indicators of health, and the incidence of clinical events such as diagnosis, recurrence and death. Focusing initially on a single biomarker and a single clinical event, the joint model methodology has been extended to take into account more complex data structures. For example, instead of one clinical event, recurrent events \cite{han2007} or multiple events have been studied either in a cause-specific context for competing risks \cite{williamson2008,elashoff2008,li2010,huang2011,li2012,Yu2010,Deslandes2010}, in a multiple correlated event context \cite{chi2006} or in a recurrent events and terminal event context \cite{liu2009,kim2012}.

In some pathologies, in addition to the multiplicity of clinical events, multiple biomarkers or outcomes are also collected over time. This is usually the case when the longitudinal process of interest is hard to capture through observed outcomes. For example, in Alzheimer's disease, cognitive functioning is measured by multiple psychometric tests in order to better approximate the underlying cognition \cite{proust2006}. In other contexts, quality of life is measured by a series of items to capture entirely the underlying concept, or functional dependency is defined by a series of activities of daily living.

In medical research, these very rich longitudinal data have long been summarized using arbitrary sum-scores or Z-scores, albeit at the price of a loss of information and rough approximation of the data. However, in psychometric research, linking observed outcomes to underlying concepts has long been explored using the latent variable models including structural equation models for continuous outcomes \cite{muthen2002_SEM} and item response models for binary and ordinal outcomes \cite{baker2004}. In biostatistics, such models also emerged more recently \cite{sammel1996,dunson2000} and were then extended to model the underlying latent process or latent trait of a series of longitudinal markers \cite{roy2000,proust2006,dunson2003}
. Since markers such as cognitive measures, \change{functional dependency scales} or quality of life questionnaires involve peculiar types of data mixing binary, ordinal, discrete and continuous (but rarely Gaussian) outcomes, Proust-Lima et al. \cite{proust-lima2012} provided a general latent process framework that simultaneously models multiple repeated outcomes of different nature, especially non-Gaussian quantitative outcomes and ordinal outcomes, that were generated by the same underlying latent trait.

The present paper combines this latent process mixed model with a joint model approach for competing risks in order to describe the natural history of cognitive functioning in the elderly -as measured by a series of psychometric tests- in association with dementia and also taking into account the competing death and the delayed entry in the cohort. \change{Because dementia is a pathology associated with old age and since dementia and death have many common risk factors, it is important to take into account the competing death when studying and predicting the risk of dementia}.

\change{Joint models mostly consist in linking the longitudinal and survival processes through shared latent variables. The random effects from the mixed models are commonly shared with the survival models by including a function of them as covariate \cite{rizopoulos2012}. Although very appealing, this approach assumes that the form of the dependency between the two types of processes is known a priori, that the same random variables model both the correlation between the repeated measures of the longitudinal markers and the correlation with the clinical events, and that the population is homogeneous}. 

\change{In this work, we preferred to link the cognitive trajectory and the risks of death and dementia using shared latent classes. These latent classes formalize latent sub-populations that are characterized by specific profiles of change for the marker and of risk for the events. As such, in addition to separating the two types of correlation and flexibly modelling the dependency, they also model an expected heterogeneity in the longitudinal trajectories and risks of event \cite{proust-lima2014}.}

\change{Joint models rely on the assumption of independence between the longitudinal and survival processes given the shared variables which needs to be checked. In the joint latent class model, it means that the structure of latent groups is supposed to capture entirely} the correlation between the longitudinal process and the risks of event. We properly assess this assumption by extending a score test previously proposed \cite{jacqmin-gadda2010} to the case of multivariate longitudinal and competing events.

The methodology is applied to describe the profiles of semantic memory associated with the onset of dementia (and death) in a cohort of elderly subjects followed up for 20 years in South-Western France. The model is also used to provide individual dynamic predictions of the risk of dementia based on cognitive measures. \change{Thanks to the joint modelling of cognitive measures and risk of dementia through shared latent classes, the Bayes theorem provides individual conditional predicted probabilities of event based on the repeated cognitive measures. Thanks to the modelled competing death, these predictions are also free of the usual bias due to the non-negligible risk of death in the elderly. }

Section 2 describes the joint latent class and latent process model. Section 3 details the estimation procedure and provides the methodology for assessing the model, especially the conditional independence assumption. Section 4 provides the application on the real cognitive data and Section 5 summarizes a simulation study that validates the estimation program and assesses the type-I error of the score test. Finally section 6 concludes.

\section{The joint latent class and latent process model}

\subsection{Model overview and notations}

\begin{figure}[h]
\centering
\includegraphics[width=0.5\textwidth]{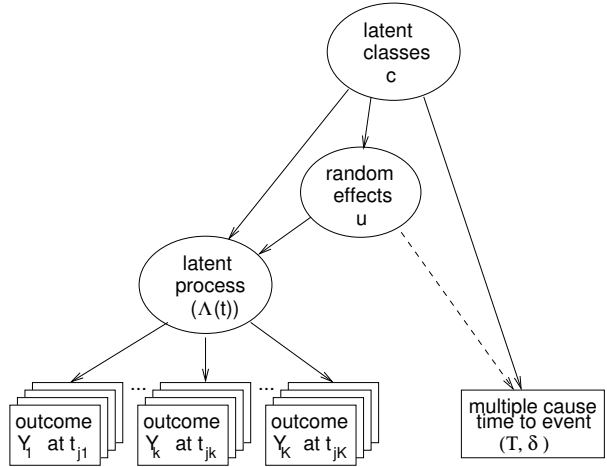}
\caption{Directed graph defining the joint latent class and latent process model for $K$ longitudinal outcomes and a multiple cause time-to-event $(T,\delta)$. The subject subscript $i$ is omitted for readability. The longitudinal outcomes $Y_k$ ($k=1,...,K$) generated by the continuous-time latent process $(\Lambda(t))_{t\in\mathbb{R}^+}$ are measured at times $t_{jk}$ ($j=1,...,n_k)$. The dashed arrow from the random effects $u$ towards the multiple cause time to event only applies in the model under the alternative hypothesis in the score test for assessing the independence assumption conditional to the latent class structure $c$.}\label{DAG}
\end{figure}

The joint model we propose is described in figure \ref{DAG}. The model relies on the existence of a structure of $G$ latent groups defined by the latent discrete variable $c$. For subject $i$ ($i=1,...,N$), $c_i=g$ if subject $i$ belongs to latent class $g$ ($g=1,...,G$). The model assumes that this latent class generated homogeneous mean profiles of the latent process of interest which is denoted $(\Lambda_i(t))_{t\in\mathbb{R}^+}$ and homogeneous risks of event. We note $T^*_{ip}$ the time to the clinical event of cause $p$ ($p=1,...,P$), and $\tilde{T}_{i}$ the time to censoring so that $T_i=\min(\tilde{T}_i,T^*_{i1},...,T^*_{iP})$ is observed with indicator $\delta_i=p$ if the clinical event of nature $p$ occurred first or $\delta_i=0$ if the subject was censored before any event. 
The time of entry in the cohort is noted $T_{i0}$. The $N$ included subjects necessarily satisfied the condition $T^*_{ip}>T_{i0}$ $\forall p$, leading to left truncation. Finally, the latent process $(\Lambda_i(t))_{t\in\mathbb{R}^+}$ generated a set of $K$ observed repeated markers. The marker $k$ ($k=1,...,K$) measured on subject $i$ ($i=1,...,N$) at occasion $j$ ($j=1,...n_{ik}$) is denoted $Y_{kij}$ and the corresponding time of measurement is denoted $t_{kij}$. Note that the repeated measures may be collected at varying times between subjects and between markers. Each part of the model is described in detail below.

\subsection{Class-specific latent process trajectory: a structural linear mixed model}

The model assumes that within each latent class $g$, the trajectory of the latent process for each subject $i$ is explained by a set of covariates, time and individual random effects using a linear mixed model:

\begin{equation}\label{LMM}
\Lambda_{i}(t)|_{c_i=g}=X_{\Lambda i}(t)^{\top} \beta_g +Z_i(t)^{\top} u_{i(g)} + w_i(t), ~~~~ \forall t \in \mathbb{R}^+
\end{equation}

where $X_{\Lambda i}(t)$ and $Z_{i}(t)$ are distinct vectors of covariates that possibly depend on time $t$; $Z_{i}(t)$ will generally include a finite basis of functions of time that captures the shape of the trajectory with time. The vector $\beta_g$ includes the association parameters of $X_{\Lambda i}(t)$  with the latent process which may or may not be specific to the latent classes depending on the clinical assumption. The r-vector of random effects $u_i$ that captures the individual variability in the trajectories has a class-specific distribution $u_{i(g)}= u_i|_{c_i=g} \sim \mathcal{N}(\mu_g,B_g)$ where the expectation $\mu_g$ will typically provide the mean profile of trajectory with time and $B_g$ will define the correlation structure between the random effects. For identifiability, we reduce to $B_g=\omega_g^2B$ with $B$ an unstructured matrix of covariances and $\omega_G=1$ with $G$ the reference class. This model does not involve any measurement error as $\Lambda_{i}$ is not observed. However, it includes a zero-mean Gaussian auto-correlated process $(w_i(t))_{t \in \mathbb{R}^+}$ for more flexibility, with for example $\text{cov}(w_i(s),w_i(t))=\sigma_w^2\min(s,t)$ for a Brownian motion or $\text{cov}(w_i(s),w_i(t))=\sigma_{w1}^2\exp(\sigma_{w2} |s-t|)$ for an auto-regressive process.

\subsection{\change{Marker-specific link with the latent process: nonlinear measurement models}}

Each observation of a longitudinal marker is linked to the latent process by a marker-specific nonlinear measurement model. This measurement model is composed of two parts: (1) a linear model links the latent process $\Lambda_i(t_{kij})$ to an intermediate noisy continuous variable at each time of measurement $\tilde{Y}_{kij}$, and (2) a transformation links the observed marker $Y_{kij}$ to its underlying intermediate noisy continuous variable $\tilde{Y}_{kij}$. From this, the nonlinear measurement model can be defined as:
\begin{equation}\label{eqobs}
\begin{split}
&Y_{kij}|_{c_i=g}=H_k(\tilde{Y}_{kij}|_{c_i=g};\eta_k)\\
\text{and  } &\tilde{Y}_{kij}|_{c_i=g}  = \Lambda_i(t_{kij}|_{c_i=g})+X_{Yi}(t_{kij})^{\top}\gamma_k + \upsilon_{ki}+\epsilon_{kij}
\end{split}
\end{equation}
where $X_{Yi}(t_{kij})$ is a vector of covariates differently associated to the observed outcomes through a vector of contrast parameters $\gamma_k$ \change{ with $\sum_{k=1}^K \gamma_k=0$ to ensure identifiability in case the same covariate has an effect on the latent process and marker-specific effects. The scalar} $\upsilon_{ki}$ is a random intercept that can capture the marker-specific inter-individual additional variability ($\upsilon_{ki} \sim \mathcal{N}(0,\sigma_{\upsilon_k}^2)$) and $\epsilon_{kij}$ is the independent measurement error ($\epsilon_{kij} \sim \mathcal{N}(0,\sigma_{\epsilon_k}^2)$). \change{Note that these marker-specific random deviations can be easily extended to account for a marker-specific random slope, for instance.} Finally $H_k$ is a link function that transforms the intermediate variable into the observed score \cite{proust2006,proust-lima2012}:
\begin{itemize}
\item For quantitative markers, $H_k^{-1}$ are defined as monotonic increasing continuous functions that depend on a vector of parameters $\eta_k$. A linear transformation reduces to the Gaussian framework but nonlinear transformations open up to continuous or discrete outcomes that are not necessarily Gaussian. For example, cumulative distribution functions of Beta distribution and I-spline bases constitute very parsimonious and flexible link functions \cite{proust-lima2012}.

\item For ordinal or binary variables, equation \eqref{eqobs} can reduce to a probit model with $H_k(\tilde{y};\eta_k)=l$ if $\tilde{y} \in [\eta_{kl},\eta_{k(l+1)}]$ (or equivalently a logistic model if $\epsilon_{kij}$ followed a logistic distribution).

\item The model also applies to bounded (or censored) quantitative markers in [min$_k$,max$_k$] by mixing the two previous definitions: $H_k^{-1}$ defined as a parametrized monotonic increasing function in (min,max), $H_k(\tilde{y};\eta_k)=\min_k$ if $\tilde{y} \leq H_k^{-1}(\min_k)$,  $H_k(\tilde{y};\eta_k)=\max_k$ if $\tilde{y} \geq H_k^{-1}(\max_k)$.

\item The model also applies to other data such as counts by adapting the $H_k$ definition.
\end{itemize}

\change{The key assumption of this multivariate longitudinal model is that the latent process captures the whole correlation between the repeated observations of two distinct markers through the correlated random effects $u_i|_{c_i=g}$ and the autocorrelated process $(w_i(t))$. Repeated measures of the same marker can have an additional correlation through the marker-specific random intercept $\upsilon_{ki}$.}

\subsection{Class-specific risks of events: a proportional hazard model}

The model assumes that the risk of event is homogeneous within each class and is explained according to covariates. We focus on proportional hazard models but the methodology also applies in other contexts. The risk of event of cause $p$ in class $g$ for subject $i$ is:
\begin{equation}\label{PHM}
\alpha_{ip}(t)|_{c_i=g}=\alpha_{0pg}(t;\nu_{pg})\exp(X_{Ti}^{\top} \zeta_{pg}) , ~~~~ \forall t \in \mathbb{R}^+
\end{equation}
where $X_{Ti}$ is a vector of covariates associated to the risk of event of cause $p$ in class $g$ by the vector of parameters $\zeta_{pg}$. The baseline risk function $\alpha_{0pg}$ is specific to each cause and each latent class. It is modelled parametrically (with $\nu_{pg}$)  using a standard survival distribution such as Gompertz or Weibull or by using a small number of step functions or M-splines for more flexibility.

\subsection{Latent class membership: a multinomial logistic regression}

Each subject $i$ belongs to a single latent class $g$. Class membership probability can be described by a multinomial logistic regression model according to covariates:
\begin{equation}\label{proba}
P(c_i=g)=\pi_{ig}=\dfrac{\exp(X_{ci}^{\top}\xi_{g})}{\sum_{l=1}^G \exp(X_{ci}^{\top}\xi_{l})}
\end{equation}
where $X_{ci}$ is a vector of covariates associated with the vector of class-$g$-specific parameters $\xi_{g}$. $X_{ci}$ includes the intercept and can reduce to it when no predictor of class membership is assumed in practice. A reference class is required for identifiability. We choose class $G$ so that $\xi_{G}=0$.

\section{Estimation and assessment}

\subsection{Maximum likelihood estimators}\label{MLE}

For a given number of latent classes $G$, the total vector of parameters $\theta_G$ consists of all the $\text{np}_G$ parameters $(\beta_1,...,\beta_G,\mu_1,...,\mu_G,\text{vec}(B),\omega_2,...,\omega_G,\sigma_w)$, $(\nu_{p1},\zeta_{p1},...,\nu_{pG},\zeta_{pG})_{p=1,...,P}$, $(\xi_1,...\xi_G)$ and \\
$(\eta_1,...,\eta_K,\gamma_1,...,\gamma_{K-1},\sigma_{\upsilon_1},...,\sigma_{\upsilon_K},\sigma_{\epsilon_1},...,\sigma_{\epsilon_K})$ involved respectively in equations \eqref{LMM}, \eqref{eqobs}, \eqref{PHM} and \eqref{proba}. The individual contribution to the log-likelihood is split using the conditional independence assumption shown in Figure \ref{DAG} (in the absence of the dashed arrow):
\begin{equation}\label{loglik}
L_i(\theta_G)=\sum_{g=1}^{G} f(Y_i \mid c_i = g;\theta_G)f(T_i,\delta_i \mid c_i = g;\theta_G)P(c_i=g\mid \theta_G)
\end{equation}
where $P(c_i=g\mid \theta_G)=\pi_{ig}$ is defined in \eqref{proba} and the density of the multiple cause time-to-event is:
\begin{equation}\label{liksurv}
f(T_i,\delta_i \mid c_i = g;\theta_G)=e^{-\sum_{p=1}^P A_p(T_i \mid c_i=g;\theta_G)}\prod_{p=1}^P \alpha_p(T_i \mid c_i=g;\theta_G)^{1_{\delta_i=p}}
\end{equation}
with $\alpha_p(t \mid c_i=g;\theta_G)$ the cause-$p$-specific instantaneous hazard defined in \eqref{PHM} and $A_p(t \mid c_i=g;\theta_G)$ the corresponding cumulative hazard.

For $f(Y_i \mid c_i = g;\theta_G)$, the density of $Y_i=(Y_{1i1},...,Y_{1in_{i1}},...,Y_{Ki1},...,Y_{Kin_{iK}})$, two cases are distinguished:
\begin{itemize}
\item Only continuous transformations are assumed for the $K$ markers. In that case, using the Jacobian of the transformations $H_k^{-1}$ for $k=1,...,K$, a closed form of the density is given by:
\begin{equation}
f(Y_i \mid c_i = g;\theta_G) = f(\tilde{Y}_i \mid c_i = g;\theta_G) \times \prod_{k=1}^K \prod_{j=1}^{n_{ik}} J(H_k^{-1}(Y_{kij}))
\end{equation}
where $J$ denotes the Jacobian and $f(. \mid c_i = g;\theta_G)$ is the density function of a multivariate normal variable with mean $E_{i,g}=(E_{i1,g}^{\top},...,E_{iK,g}^{\top})^{\top}$ and covariance matrix $V_{i,g} = Z_i B_g Z_i^{\top} + R_i + \Sigma_i$ where $E_{ik,g}= X_{\Lambda ik} \beta_g +Z_{ik} \mu_g + X_{Yik} \gamma_k$. The design matrices $X_{\Lambda ik}$, $Z_{ik}$ and $X_{Yik}$ have row vectors $X_{\Lambda i}(t_{kij})$, $Z_{i}(t_{kij})$ and $X_{Yi}(t_{kij})$ for $j=1,...n_{ik}$, $Z_{i}=(Z_{i1}^{\top},...,Z_{iK}^{\top})^{\top}$, $R_i$ defines the covariance matrix of the auto-correlated process $(w_i(t))_{t \in \mathbb{R}^+}$, and $\Sigma_{i}$ is the $K$-bloc diagonal matrix with $k^{\text{th}}$ block $\Sigma_{ik}=\sigma_{\upsilon_k}^2J_{n_{ik}}+\sigma_{\epsilon_k}^2I_{n_{ik}}$, $J_n$ the $n \times n$- matrix of elements 1, and $I_{n}$ the $n \times n$- identity matrix.

 \item At least one transformation is defined as a threshold transformation or a transformation for bounded outcome. Let $K_1$ be the number of outcomes with a threshold transformation or a transformation for bounded outcome ($K_1 \in \{1,...,K\}$). In that case, no Gaussian process $(w_i(t))_{t \in \mathbb{R}^+})$ is considered and the density is decomposed according to the individual random effects $u_i$ and $\upsilon_{ki}$:
\begin{equation} \label{condY}
\begin{split}
f(Y_i \mid c_i = g;\theta_G) = &\displaystyle \int ~ \left ( \prod_{k=1}^{K_1}  ~  \int \prod_{j=1}^{n_{ik}} f_y(Y_{kij} | c_i = g,u_{i(g)}=u,\upsilon_{ki}=\upsilon) ~ f_{\upsilon_g}(\upsilon|c_i = g)d\upsilon~
\right ) \\
&\left ( \prod_{k=K_1+1}^{K} f(\tilde{Y}_{ki} \mid c_i = g,u_{i(g)}=u;\theta_G) \times \prod_{j=1}^{n_{ik}} J(H_k^{-1}(Y_{kij})) \right ) f_{u_g}(u|c_i = g)d{u}
\end{split}
\end{equation}
where $f_{u_g}$ and $f_{\upsilon_g}$ are the density functions of a multivariate normal with mean $\mu_g$ and covariance matrix $B_g$, and of a Gaussian distribution with mean 0 and variance $\sigma_{\upsilon_k}^2$. For $k>K_1$, $f(\tilde{Y}_{ki} \mid c_i = g,u_{i(g)}=u;\theta_G)$ is the density function of a multivariate normal variable with mean $X_{\Lambda ik} \beta_g +Z_{ik} u_{i(g)} + X_{Yik} \gamma_k$ and covariance matrix $\Sigma_{ik}$. For $k \leq K_1$, $f_y(Y_{kij} | c_i = g,u_{i(g)}=u,\upsilon_{ki}=\upsilon)$ is defined in the appendix.
\end{itemize}

To take into account the delayed entry in the cohort, the log-likelihood for left-truncated data $l^{T_0}(\theta_G)=\sum_{i=1}^N \log \left (\dfrac{L_i(\theta_G)}{S_i(T_{i0};\theta_G)} \right )$ was considered instead of the standard log-likelihood $l(\theta_G)=\sum_{i=1}^N \log \left (L_i(\theta_G) \right )$ with $S_i(T_{i0};\theta_G)=\sum_{g=1}^G \pi_{ig} e^{-\sum_{p=1}^P A_p(T_{i0} \mid c_i=g;\theta_G)}$ the marginal survival function in $T_{i0}$. The log-likelihood $l^{T_0}$ was maximized using a \change{robust Newton-like algorithm, the} Marquardt algorithm, for a fixed number of latent classes. The algorithm ensures that the log-likelihood is improved at each iteration. Gradients and the Hessian matrix were computed by finite differences. The estimation procedure was implemented in HETMIXSURV program (Version 2 - \url{http://www.isped.u-bordeaux.fr/BIOSTAT}), a Fortran90 parallel program in which the convergence is based on three simultaneous criteria at iteration $i$ on: (1) the parameter estimates ($\sum_{l=1}^{\text{np}_G} |\theta_{Gl}^{(i)}-\theta_{Gl}^{(i-1)}|^2 < \epsilon_1$), (2) the log-likelihood ($|l(\theta_{G}^{(i)})-l(\theta_{G}^{(i-1)})| < \epsilon_2$), (3) the first and second derivatives $\mathcal{G}^{(i)}$ and $H^{(i)}$ ($\frac{\mathcal{G}^{(i)\top} H^{(i)-1}\mathcal{G}^{(i)}}{\text{np}_G}< \epsilon_3$), $\epsilon_\text{.}<$0.0001 in the application. In practice, the algorithm is run from multiple sets of initial values to ensure the convergence towards the global maximum as it is recommended in mixture models \cite{hipp2006}. \change{The optimal number of latent classes can be selected from a series of criteria. In this work, we reported the Bayesian Information Criterion (BIC=$-2l^{T_0}(\hat{\theta}_G)+\text{np}_G \text{log}(N)$) and a score test statistic presented in section \ref{scoretest} that assesses the conditional independence assumption. In case of discordance between criteria, we favored the latter to select the optimal number of latent classes.}

\change{Inference was based on the asymptotic normality of the parameters with the variance-covariance of $\hat{\theta}_G$ estimated by the inverse of the Hessian matrix at the optimum.} \change{For some meaningful functions of the parameters for which no direct inference could be obtained (individual dynamic predictions of event or predictions in the natural scale of the markers), a Monte-Carlo method was also used. It consisted in approximating the posterior distribution of the function from a large set of draws from the asymptotic normal distribution of the parameters  \cite{proust-lima2009biostat}. This method is also called parametric bootstrap.} 

\subsection{Posterior probabilities}\label{postprob}

Posterior probabilities can be computed from the joint model. Posterior class-membership probabilities given all the \change{observations for the markers and the events} 
 $\pi_{ig}^{(Y,T)}(\theta_G)=P(c_i=g|Y_i,(T_i,\delta_i);\theta_G)$ are used to build a posterior classification of the subjects with posterior affectation given by $\hat{c_i}=\text{argmax}(\pi_{ig}^{(Y,T)}(\hat{\theta}_G))$. They are also used to evaluate the discrimination of the latent classes as illustrated in section \ref{gof}.

Joint models can also provide individual dynamic probabilities of having the event in a window of time $[s,s+t]$ given the biomarker history up to the time of prediction $s$. This was extensively described in the standard joint model setting with one longitudinal marker and one event \cite{proust-lima2009biostat,rizopoulos2012,proust-lima2014}. In a multivariate setting with multiple outcomes and competing risks, individual predictions can also be computed using the posterior cumulative incidence of each cause of event in the window of times $[s,s+t]$ given the outcomes history $Y_i ^{(s)}=\{Y_{kij},\, k=1,...,K, j=1,..., n_{ik}, \text{such as}\; t_{kij}\leq s\}$, the covariates history $X_{i}^{(s)}=\{X_{\Lambda i}(t_{kij}),Z_{i}(t_{kij}),X_{Yi}(t_{kij}) ,\, k=1,...,K, j=1,..., n_{ik}, \text{such as}\; t_{kij}\leq s\}$ and the time-independent covariates $X_{Ti}$ and $X_{ci}$. It is defined for cause $p$ as:
\begin{equation}
\begin{split}
P_{pi}(s,t)&=P(T_i \leq s+t, \delta_i=p | T_i>s,Y_i^{(s)},X_{i}^{(s)},X_{Ti},X_{ci};\theta_G)\\
&=\frac{\sum_{g=1}^G P(c_i=g|X_{ci};\theta_G)P(T_i \in (s,s+t],\delta_i=p | X_{Ti},c_i=g;\theta_G)f(Y_i^{(s)}|X_{i}^{(s)},c_i=g ;\theta_G)}
{\sum_{g=1}^G P(c_i=g|X_{ci};\theta_G)S_i(s| X_{Ti},c_i=g;\theta_G)f(Y_i^{(s)}|X_{i}^{(s)},c_i=g ;\theta_G)}
\end{split}
\end{equation}

in which the class-specific cause-specific cumulative incidence
\begin{equation}\label{cuminc}
P(T_i \in (s,s+t],\delta_i=p | X_{Ti},c_i=g;\theta_G)=\int_s^{s+t} \alpha_p(u \mid c_i=g;\theta_G) \exp \left ( - \sum_{l=1}^P A_l(u \mid c_i=g;\theta_G) \right)
\end{equation}
and the density of the longitudinal outcomes $f(Y_i^{(s)}|X_{i}^{(s)},c_i=g ;\theta_G)$ in class $g$, the class-specific membership probability $P(c_i=g|X_{ci};\theta_G)$, the class-specific survival function $S_i(s| X_{Ti},c_i=g;\theta_G)$, the cause-specific hazards $\alpha_p(u \mid c_i=g;\theta_G)$ and cumulative hazards $A_p(u \mid c_i=g;\theta_G)$ are all defined similarly as in section \ref{MLE}. This cause-specific cumulative incidence requires the numerical computation of the integral. This is achieved by a 50-point Gauss-Legendre quadrature.

\subsection{Assessment of the conditional independence assumption}\label{scoretest}

Joint latent class models rely on the conditional independence assumption of the longitudinal outcomes and the times-to-event given the latent classes. We evaluated this assumption by extending the score test proposed by Jacqmin-Gadda et al. \cite{jacqmin-gadda2010} to the multivariate longitudinal setting, the competing event setting and the left-truncation setting. The test evaluates whether, conditional on the latent classes, there exists a residual dependency between the longitudinal and the survival processes through the random effects, as shown in figure \ref{DAG} with the dashed arrow. Under the alternative assumption $\mathcal{H}_1$, the cause-specific survival model defined in \eqref{PHM} for cause $p$ ($p=1,...,P$) becomes
\begin{equation}\label{PHM_H1}
\alpha_{p}^{\mathcal{H}_1}(t)|_{c_i=g,u_{i(g)}}=\alpha_{0pg}(t;\nu_{pg})\exp(X_{Ti}^{\top} \zeta_{pg}+u_{i(g)}^*\kappa_p) , ~~~~ \forall t \in \mathbb{R}^+
\end{equation}
where $\kappa_p$ is the vector of parameters associating the centered random effects $u_{i(g)}^*=u_{i(g)}-E(u_{i(g)})$ with cause $p$ of event.

The log-likelihood for truncated data becomes $l^{T_0,\mathcal{H}_1}(\theta_G)=\sum_{i=1}^N \log \left (\dfrac{L_i^{\mathcal{H}_1}(\theta_G)}{S_i^{\mathcal{H}_1}(T_{i0};\theta_G)} \right )$ where $S_i^{\mathcal{H}_1}(T_{i0};\theta_G)=\sum_{g=1}^G \pi_{ig} e^{-\sum_{p=1}^P A_p^{\mathcal{H}_1}(T_{i0} \mid c_i=g;\theta_G)}$ is the marginal survival function in $T_{i0}$ under $\mathcal{H}_1$, and the individual contribution to the likelihood is
\begin{equation}\label{loglik_H1}
L_i^{\mathcal{H}_1}(\theta_G)=\sum_{g=1}^{G} P(c_i=g\mid \theta_G) \int f(Y_i \mid c_i = g,u_{i(g)}=u;\theta_G)f(T_i \mid c_i = g,u_{i(g)}=u;\theta_G)f(u\mid c_i = g;\theta_G)du
\end{equation}

The cause-specific score test consists in testing $\mathcal{H}_0:\kappa_p=0$ (\emph{versus} $\mathcal{H}_1:\kappa_p \not= 0$) for cause $p$. After lengthy calculation similar to Jacqmin-Gadda et al. \cite{jacqmin-gadda2010}, we demonstrate that the score test statistic equals:
\begin{equation}
\begin{split}
U_p&=\sum_{i=1}^N \dfrac{\partial \log(L_i^{\mathcal{H}_1}(\theta_G))}{\partial \kappa_p}|_{\kappa_p=0,\hat{\theta}_G} -  \sum_{i=1}^N \dfrac{\partial \log(S_i^{\mathcal{H}_1}(T_{i0};\theta_G))}{\partial \kappa_p}|_{\kappa_p=0,\hat{\theta}_G} \\
 &=\sum_{i=1}^N \sum_{g=1}^G P(c_i=g | Y_i,(T_i,\delta_i); \hat{\theta}_G)\left (1_{\delta_i=p}-A_p(T_{i}|c_i=g;\hat{\theta}_G)\right)E(u_i^*|c_i=g,Y_i;\hat{\theta}_G)
\end{split}
\end{equation}
The conditional expectation $E(u_i^*|c_i=g,Y_i;\hat{\theta}_G)$ is replaced by the empirical Bayes estimate when only continuous outcomes are modelled or computed by numerical integration (Gauss-Hermite quadrature) otherwise. The right part of $U$ corresponding to the left-truncation in $T_{0i}$ reduces to 0.

The global score test can also be computed by testing $\mathcal{H}_0:\kappa=(\kappa_1^{\top},...,\kappa_P^{\top})^{\top}=0$ (\emph{versus} $\mathcal{H}_1:\kappa \not= 0$) with the statistic $U=(U_1^{\top},...U_p^{\top})^{\top}$. Under the null hypothesis, $U^{\top}\text{Var}(U)^{-1} U$ and $U_p^{\top}\text{Var}(U_p)^{-1} U_p$ follow a chi-square with $r\times P$ and $r$ degrees of freedom, respectively. The variances $\text{Var}(U_p)$ and $\text{Var}(U)$ are approximated by the empirical variances of the individual contributions to $U_p$ and $U$ respectively \cite{jacqmin-gadda2010}.

\section{Application: profiles of semantic memory in the elderly}

Along with episodic memory considered as the hallmark of Alzheimer's disease (AD), semantic memory, which refers to the memory of meanings and other concept-based knowledge, could play an important role in the development of AD and dementia. Indeed, it seems to be affected very early in the prodromal phase of AD with a decline appearing as early as 12 years before the diagnosis of dementia \cite{amieva2008}. In this context, the aim of this application was to describe profiles of semantic memory in the elderly in association with the competing risks of dementia and death using the proposed methodology.

\subsection{Sample from Paquid prospective cohort}

Data came from the French prospective cohort study PAQUID initiated in 1988 in southwestern France to explore functional and cerebral aging \cite{jacqmin1997} 
. In brief, 3777 subjects who were older than 65 years were randomly selected from electoral rolls and were eligible to participate if they were living at home at the time of enrollment. The subjects were extensively interviewed at home by trained psychologists at baseline (V0) and were followed up at years 1, 3, 5, 8, 10, 13, and 15, 17, 20 and 22. At each visit, a battery of psychometric tests was completed and a two-phase screening procedure for the diagnosis of dementia was carried out. We considered here two tests of semantic memory: the Isaacs Set test (\texttt{IST15}) \cite{isaacs1973} shortened to 15 seconds and the Wechsler Similarities test (\texttt{WST}) \cite{wechsler1981}. In \texttt{IST15}, subjects are required to name words (with a maximum of 10) belonging to a specific semantic category (cities, fruits, animals and colors) in 15 seconds. The score ranges from 0 to 40. The \texttt{WST} consists in saying for a series of five pairs of words to what extent the two items are alike. The score ranges from 0 to 10. For both tests, low values indicate a more severe impairment.

 We focused in the analysis on the subset of subjects for whom the main genetic factor associated with cognitive aging, ApoE4 for Apolipoprotein E4, was available. We did not consider cognitive measurements at the initial visit because of a learning effect previously described \cite{jacqmin1997} between the first two exams, but included all the subjects with at least one measure at each test during the 21-year follow-up and free of dementia at 1 year. Death free of dementia was defined as any death in the two years following a negative diagnosis of dementia. This prevented us from the interval censoring issue with dementia diagnosis. In addition to ApoE4 carriers, we considered two covariates: gender and educational level in two classes (subjects who graduated from primary school and those who did not).

The sample consisted of 588 subjects including 333 (56.6\%) women, 436 (74.1\%) who graduated from primary school and 127 (21.6\%) ApoE4 carriers. The subjects had a median of 3 measures on the \texttt{IST15} (IQR=2-6) and 4 measures on the \texttt{WST} (IQR=2-7). \texttt{IST15} measures at the 3-year visit were excluded since a subsample of PAQUID completed a nutritional questionnaire at V3 that may have impacted the fruit and animal fluency subscores. The two test distributions are given in the supplementary material (Figure S1 - top). The \texttt{WST} distribution is relatively atypical with a large proportion (30.0\%) of \texttt{WST} scores at the maximum value of 10. For the \texttt{IST15}, the distribution is also asymmetric with more values in high scores but remains closer to the normal distribution. Dementia was considered as a terminating event so measures collected after diagnosis were not included in the analyses, and age at dementia was defined as the mean age between the age at the diagnosis visit and the age at the preceding visit. A total of 171 (29.1\%) subjects had incident dementia and 245 (41.7\%) died free of dementia.

\subsection{\change{Step-by-step construction of the model}}

\change{Owing to its complexity and the many parametric options, the joint latent class and latent process model was built progressively according to the following steps: }

\begin{itemize}

\item \change{First, appropriate link functions were selected. Each longitudinal marker was modelled separately in a latent process mixed model (without latent classes) in which the underlying latent process trajectory was a quadratic function of age with 3 correlated random effects and no adjustment for covariates:
\begin{equation}
\Lambda_i(\texttt{age}) =  u_{0i}  + u_{1i} \texttt{age} + u_{2i} \frac{\texttt{age}^2}{10} ~~~~~~~~~\text{\&} ~~~~~~~~
\texttt{Y}_{ij}=H(\Lambda(\texttt{age}_{ij}) + \epsilon_{ij},\eta)
\end{equation}
where $\texttt{age}_{ij}$ is the age in decades from 65 years old for subject $i$ at occasion $j$, $u_{i} \sim \mathcal{N}(\mu,B)$ with $B$ unstructured, $\epsilon_{ij} \sim \mathcal{N}(\sigma_{\epsilon}^2)$ and $\texttt{Y}_{ij}$ was successively $\texttt{IST15}_{ij}$ and $\texttt{WST}_{ij}$. }

\change{Different families of link functions $H(.,\eta)$ were compared: the linear, the splines and the threshold link functions. For the approximation with splines, we considered 3 internal knots as a compromise between flexibility and a parcimonious number of parameters, and placed the knots around the quantiles (23,27,31 for \texttt{IST15} and 6,8,9 for \texttt{WST}). The most appropriate link function was selected as a balance between goodness-of-fit (as measured by the discretized AIC \cite{proust-lima2012}) and complexity of estimation. The estimated transformations and discretized AIC are provided in supplementary material (figure S1 -bottom). For \texttt{WST}, although the threshold link function gave a better discretized AIC (by respectively 38.9 and 329 points compared to the splines and the linear transformation), we retained the splines transformation for the rest of the analysis. Indeed, it avoided the numerical integration over the 3 random effects (with the threshold link function) while providing a reasonable fit compared to the linear link function. For \texttt{IST}, the splines link function was selected since it provided the best discretized AIC.}

\item \change{The second step consisted in building the bivariate latent process mixed model for the two longitudinal markers by assuming they were two measures of the same latent semantic memory process $\Lambda$:
\begin{equation}
\begin{split}
\Lambda_i(\texttt{age}) &=  u_{0i}  + u_{1i} \texttt{age} + u_{2i} \texttt{age}^2\\
\texttt{IST15}_{ij}&=H_1(\Lambda(\texttt{age}_{1ij}) + \upsilon_{1i} +  \epsilon_{1ij},\eta_1) ~~~~~~~~~\text{\&} ~~~~~~~~
\texttt{WST}_{ij} =H_2(\Lambda(\texttt{age}_{2ij})+ \upsilon_{2i} + \epsilon_{2ij},\eta_2) \\
\end{split}
\end{equation}
where $\upsilon_{li} \sim \mathcal{N}(0,\sigma_{\upsilon_l}^2)$ and $\epsilon_{lij} \sim \mathcal{N}(0,\sigma_l²)$ for $l=1,2$, and $H_.$ were approximated by I-splines with 3 internal knots as in the first step. Considering this bivariate model compared to the two separate models improved the fit by roughly 200 points of AIC while using 5 fewer parameters. As $\sigma_{\upsilon_1}$ was estimated at 0, $\upsilon_{1i}$ was set to 0 for the rest of the analysis. Finally, a Brownian motion and adjustment for covariates in $\Lambda(\texttt{age})$ were considered as formulated below in equation \eqref{joint_appli_final}.}

\item \change{The third step consisted in specifying the cause-specific baseline risk functions. The risks of dementia and dementia-free death were first modelled in a standard cause-specific model (without latent classes) according to age and adjusted for gender (noted \texttt{sex} with woman in reference), a binary indicator of educational level (noted \texttt{EL} with subjects who did not graduate from primary school in reference) and ApoE4 carriers (noted \texttt{E4} with ApoE4 non-carriers in reference):
\begin{equation}\label{survappli}
\alpha_p(t) = \alpha_{0p}(t;)e^{\zeta_{1p} \texttt{sex} +  \zeta_{2p} \texttt{EL} +   \zeta_{3p} \texttt{E4}} \text{ , } p=1,2
\end{equation}
Different parametric shapes for $\alpha_{0p}(t)$, $p=1,2$ were compared. The 2-parameter Weibull cause-specific baseline risk functions and the 5-parameter cause-specific baseline risk functions approximated by M-splines with 3-knots \cite{proust-lima2009} provided the same AIC (AIC=3436.1 for Weibull and AIC=3436.0 for I-splines). In contrast, the model with 2-parameter Gompertz cause-specific baseline risk functions gave a largely worse fit with (AIC=3599.0). Hence, class-specific and cause-specific Weibull baseline risks functions were considered for the remainder of the paper as they provided the same fit as I-splines baseline risks while being more parcimonious.}
\end{itemize}

\subsection{\change{Final specification of the joint latent class and latent process model}}

\change{In view of the preliminary analyses, the following joint latent class model was finally considered. The class-specific trajectory of semantic memory according to age was defined as a subject-specific quadratic function of age from 65 years old, adjusted for age at entry (noted \texttt{ageT0} centered at 65years old), \texttt{sex}, \texttt{EL} and \texttt{E4}:
\begin{equation}
\begin{split}\label{joint_appli_final}
\Lambda(\texttt{age})|_{c_i=g} = &  u_{0i(g)}  + \beta_1 \texttt{ageT0} +   \beta_2 \texttt{sex} +  \beta_3 \texttt{EL} +   \beta_4 \texttt{E4} + (u_{1i(g)}  + \beta_5 \texttt{sex} +  \beta_6 \texttt{EL} +   \beta_7\texttt{E4}) \times \texttt{age} + \\
& (u_{2i(g)}  + \beta_{8} \texttt{sex} +  \beta_{9} \texttt{EL} +   \beta_{10} \texttt{E4}) \times \frac{\texttt{age}^2}{10} + w_i(\texttt{age})
\end{split}
\end{equation}
where $u_{i(g)} \sim \mathcal{N}(\mu_g,B)$ with $B$ unstructured, and $w_i(\texttt{age})$ is a Brownian motion according to age from 65 with variance $\sigma_w^2\texttt{age}$. The underlying semantic memory was related to the observed outcomes (\texttt{IST15} and \texttt{WST}) at the observed ages according to:
\begin{equation}
\texttt{IST15}_{ij}=H_1(\Lambda(\texttt{age}_{1ij}) +  \epsilon_{1ij},\eta_1) ~~~~~~\text{   \&     } ~~~~~~ \texttt{WST}_{ij}=H_2(\Lambda(\texttt{age}_{2ij})+ \upsilon_{2i} + \epsilon_{2ij},\eta_2)
\end{equation}
where $\upsilon_{2i} \sim \mathcal{N}(0,\sigma_{\upsilon_2}^2)$ and $\epsilon_{lij} \sim \mathcal{N}(0,\sigma_{\epsilon_l}²)$ for $l=1,2$, and $H_1$ and $H_2$ were approximated by splines link functions (with 3 internal knots placed respectively at 23,27,31 for \texttt{IST15} and 6,8,9 for \texttt{WST}). No contrasts of covariates were included in these equations as the contrasts were not of major interest in this application and were not significant. The class-specific risks of dementia and dementia-free death were described according to age and were adjusted for \texttt{sex}, \texttt{EL} and \texttt{E4}:
\begin{equation}\label{survappli2}
\alpha_p(t)|_{c_i=g} = \alpha_{0pg}(t;)e^{\zeta_{1p} \texttt{sex} +  \zeta_{2p} \texttt{EL} +   \zeta_{3p} \texttt{E4}} \text{ , } p=1,2
\end{equation}
with $\alpha_{0..}(t)$ the cause-and-class-specific 2-parameter Weibull baseline risk functions. Finally, the latent class membership was not modelled according to covariates so $P({c_i=g}) = \pi_g$. }

\subsection{Selection of the number of latent classes}

\begin{figure}[h]
\centering
\includegraphics[width=0.80\textwidth]{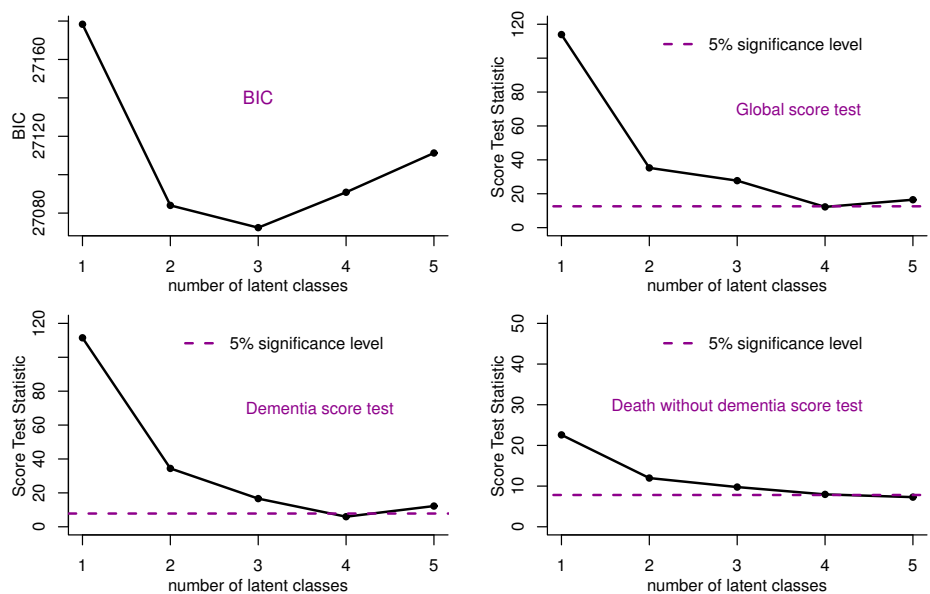}
\caption{BIC and score test statistics of the conditional independence test (global statistic and cause-specific statistics with 5\% significance levels) according to the number of latent classes.}
\label{choice_G}
\end{figure}

This joint latent class and latent process model was estimated for a number of latent classes $G$ varying from 1 to 5. \change{Each time, various sets of initial values were systematically tested. They were randomly chosen or selected by going back and forth from different $G$ and removing a class or adding a new one}. Figure \ref{choice_G} provides the BIC and the score test statistics for the global conditional independence test and the cause-specific conditional independence score tests. The BIC favored the 3-class model while the three conditional independence scores favored the 4-class model with p-values under the 5\% significance level. Based on this, we selected the 4-class model to make sure the whole dependency between the times-to-dementia-and-death and the semantic memory trajectory was captured.

\subsection{\change{Results for the final 4-class model}}

\begin{figure}[h]
\centering
\includegraphics[width=0.90\textwidth]{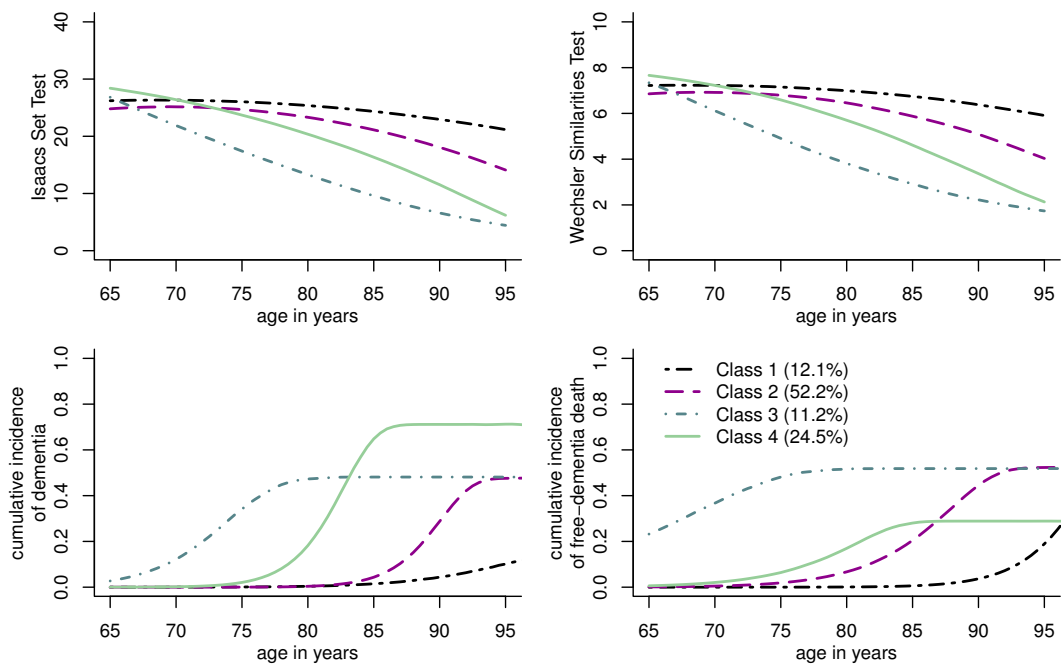}
\caption{Top: Class-specific predicted trajectory of \texttt{IST15} and \texttt{WST} for a woman, EL-, E4- who entered the cohort at 75 years. Bottom: cumulative incidence of dementia and dementia-free death adjusted for memory trajectory for a woman, EL-, E4- who entered the cohort at 75 years. The median is plotted over a 2000-draw MC approximation.}\label{evol_surv_G4}
\end{figure}

Figure \ref{evol_surv_G4} provides the 4 predicted profiles of semantic memory declines translated into each psychometric scale and the 4 cause-specific cumulative incidences of dementia and death (Figure S2 in supplementary material provides the same figure with 95\% confidence bands). \change{The parameter estimates and associated standard errors are given in Table S1 in supplementary materials.} Class 1 with 12.1\% of the sample was characterized by a slight semantic memory decline with age and close to zero risks of dementia and death before age 85. This class can be called "dementia-free survivors" with 10\% of dementia diagnosis and 25\% of death by age 95. Class 2 corresponding to the majority of the sample (52.2\%) was characterized by a progressive semantic memory decline with age that accelerated in the older ages. It was associated with an increased incidence of death from 70 years old and an increased risk of dementia after 85 years old. This class could describe "natural aging". Class 3 which corresponded to 11.2\% of the sample was characterized by a very rapid decline on semantic memory tests from the early 70s and was associated with a very high incidence of death from 65 years and a high risk of dementia in the same period. Both incidences reached a ceiling from 80 years old meaning that at 80 years, all the subjects in this class were roughly either diagnosed as demented (47.3\%) or dead without dementia (51.6\%). This class can be called "early dementia and death". Finally, class 4 with 24.5\% of the sample was intermediate with a more prononouced cognitive decline than in latent classes 1 and 2 and an increasing incidence of dementia between 75 and 88 years old, the probability of being demented before dying in this class reaching 71\% at 88 years old. This class could be called "intermediate dementia".

The four latent classes were defined after adjustment for age at entry, educational level, gender and apoE in the cognitive trajectories and the cause-specific risks of event. With these adjustments, the posterior latent classes differed significantly according to age at entry (p=0.001) in the cohort and education level (p$<$0.0001). No difference was observed according to gender (p=0.604) or apoeE4 status (p=0.095). Compared to the classes "natural aging" and "intermediate dementia" with respectively 73\% and 77\% of subjects who graduated from primary school, the "dementia-free survivors" were less likely to be highly educated subjects (only 56\%) while the "early dementia and death" were mainly highly educated subjects (90\% graduated from primary school). This latter class included mainly younger subjects (mean age at entry of 69 years old) compared to the others with mean ages at entry around 74 years old.

\begin{figure}[h]
\centering
\includegraphics[width=0.9\textwidth]{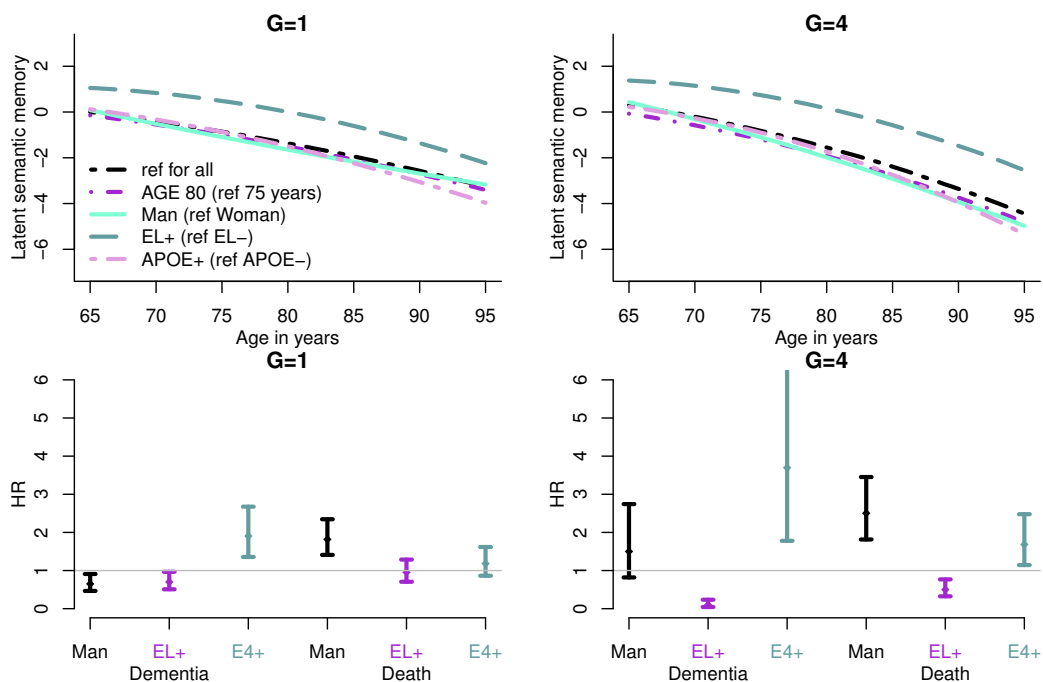}
\caption{Marginal effect of  \texttt{sex} (man compared to woman), of high educational level (\texttt{EL+} compared to low) and of ApoE4 carriers (\texttt{E4+} compared to non-carriers), and age at entry (\texttt{age0} in decades) on the mean predicted trajectory of the latent semantic memory (top) and on the cause-specific risks of events (bottom) in the 1-class model and the 4-class model. Top: mean semantic memory trajectory for a change in each covariate is plotted and compared to the trajectory in the reference category (woman ApoE4 non-carrier entered at 75 years old with a low educational level). Bottom: hazard ratios and 95\% \change{confidence intervals}.}\label{marg_covariates}
\end{figure}

The parameters in model \eqref{survappli2} show the effects of education level, apoE and gender on the risks of dementia and death, adjusted for the semantic memory decline if $G>1$. 
The corresponding hazard ratios are provided in figure \ref{marg_covariates} (bottom) both in the 4-class model and in the 1-class model. After adjustment for semantic memory, gender was no longer associated with the risk of dementia, the protective effect of educational level on dementia risk was more accentuated and apoE4 carriers had an even higher risk of dementia. Regarding death, while education and ApoE4 were not associated with the risk of death in the standard survival model, they were associated with death after adjusting for cognitive trajectory classes: highly educated subjects had a lower risk of death and ApoE+ carriers had an increased risk of death. The effect of gender with a higher risk of death for men is inflated when adjusted for the cognitive trajectory.

Trajectories of semantic memory decline are plotted in figure \ref{marg_covariates} according to covariates in both the separate longitudinal model (1-class model) and in the joint model, taking into account the informative dropout due to death and dementia (4-class model). \change{In the 4-class model that takes into account the associated times to dementia and death without dementia}, subjects with a higher educational level had a higher cognitive level at 65 years old (p=0.004) and a tendency to have a slower decline compared to subjects with a low educational level (p=0.075 for \texttt{age} and \texttt{age}$^2$). Men and women had a similar cognitive level at 65 years old (p=0.533) but men experienced a faster decline than women (p=0.050 for \texttt{age} and \texttt{age}$^2$). Similarly, cognitive level did not differ at 65 years old according to apoE4 status (p=0.737) but apoE4 carriers tended to have faster cognitive decline with age (p=0.052 for \texttt{age} and \texttt{age}$^2$). Finally, older subjects at entry had a significantly lower cognitive level whatever the time (p=0.002). When not adjusting for the dependent dropout, the covariate effects on the initial cognitive level remained unchanged but none of the interactions with time were significant (p=0.315, p=0.448, and p=0.118 for respectively \texttt{sex}, \texttt{EL} and \texttt{E4} in interaction with \texttt{age} and \texttt{age}$^2$). The major difference between the 1-class and 4-class models was the higher intensity of semantic memory decline over age when taking into account the associated dementia and death. This was expected owing to the informative dropout they constitute in the elderly.

\subsection{Goodness-of-fit evaluation}\label{gof}

\begin{figure}[h]
\centering
\includegraphics[width=0.9\textwidth]{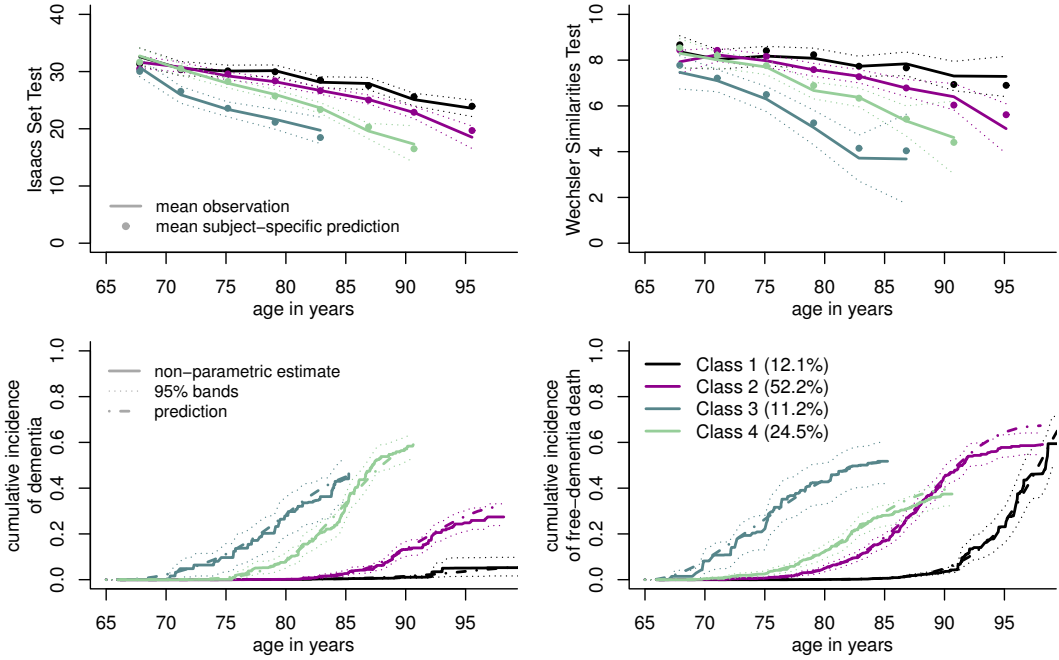}
\caption{\change{Top: Mean observed (plain lines) and mean subject-specific predicted ($\bullet$) trajectories in each latent class weighted by the posterior class-membership probabilities along with the 95\% point-wise confidence bands of the weighted mean observed trajectories (dotted lines).
 Bottom: Non-parametric Aalen-Johansen estimate (plain lines) and mean predicted (dashed-dotted lines) cumulative incidences of dementia and dementia-free death each latent class weighted by the posterior class-membership probabilities along with the 95\% point-wise confidence bands of the non-parametric estimator computed using a non-parametric bootstrap with 2000 replicates (dotted lines).}}\label{gof_eval}
\end{figure}

\change{In addition to the step-by-step construction of the model and the assessment of the conditional independence assumption, we evaluated the goodness-of-fit of the 3 submodels of the joint latent class model according to} the methodology described in Proust-Lima et al. \cite{proust-lima2014}. We \change{first} compared the subject-specific predictions to the observations in the longitudinal submodel and in the cause-specific survival model. \change{Class-specific predictions for each longitudinal marker $k$ were obtained as the weighted means of the predictions computed for a series of intervals of time $\mathcal{I}(t)$ as } 
\begin{equation}
\dfrac{\sum_{i=1}^N \sum_{j=1}^{n_{ik}} \pi_{ig}^{(Y,T)}(\hat{\theta}_G) \hat{Y}_{kijg} \mathds{1}_{t_{kij} \in \mathcal{I}(t)}}{\sum_{i=1}^N \sum_{j=1}^{n_{ik}} \pi_{ig}^{(Y,T)}(\hat{\theta}_G)\mathds{1}_{t_{kij} \in \mathcal{I}(t)}}
\end{equation}  
\change{for class $g$, $g=1,...,G$ where $\hat{Y}_{kijg}=\text{E}(H(\tilde{Y}_{kij};\hat{\eta}_k)|c_i=g,\hat{u}_{i(g)},\hat{\upsilon}_{ki},\hat{w}_i(t_{kij});\hat{\theta}_G)$ are the subject-, class-, and marker-specific predictions computed by numerical approximation \cite{proust-lima2012} with $\hat{u}_{i(g)}$, $\hat{\upsilon}_{ki}$, $\hat{w}_i(t_{kij})$ denoting the empirical Bayes estimates of the random deviations computed from the normalized data $\tilde{Y}_i$. The weighted mean longitudinal observations were obtained according to the same formula by replacing $\hat{Y}_{kijg}$ by ${Y}_{kij}$ and are displayed with their 95\% confidence bands. The class-specific cumulative incidences for cause $p$ predicted by the model were computed as the weighted means of the predicted cumulative incidences $\text{cum}_{ig}(t)=P(T_i \leq t,\delta_i=p | X_{Ti},c_i=g;\hat{\theta}_G)$ defined in \eqref{cuminc} and weighted by $\pi_{ig}^{(Y,T)}(\hat{\theta}_G)$ for class $g$, $g=1,...,G$. They were compared to the weighted Aalen-Johansen estimator obtained by R package \texttt{prodlim} \cite{prodlim}. For the latter, 95\% confidence bands were computed by non-parametric bootstrap in the absence of any other straightfoward computation technique}. These comparisons displayed in figure \ref{gof_eval} underline the very good fit of the model to the data. \\
We \change{finally} evaluated the quality of the classification obtained from the 4-class joint model using the posterior classification table in the supplementary material (Table S1). In each class, the mean posterior probability of belonging to this class ranged from 75.0\% in class 4 to 87.6\% in class 1, indicating a clear discrimination between the latent classes.

\subsection{Individual dynamic predictions}

\begin{figure}[h]
\centering
\includegraphics[width=0.6\textwidth]{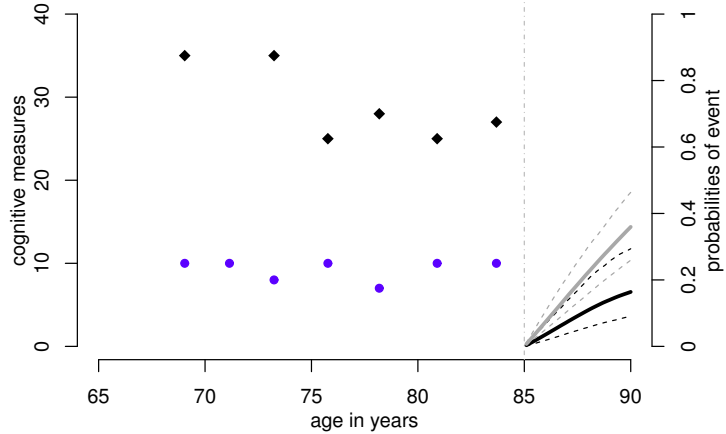}
\caption{Observed cognitive measures (IST15 in {$\Diamondblack$} and WST in {\textcolor{blue}{$\bullet$}} ) and cause-specific individual predictive cumulative incidences (of dementia in black and death in grey) up to an horizon of 5 years for a ApoE- man who graduated from primary school and entered the cohort at 68 years old. Predicted cumulative incidences are computed using a Monte-Carlo approximation with 2000 draws. Are given the median and the 2.5\% and 97.5\% percentiles.}\label{predindiv}
\end{figure}

To illustrate the individual dynamic cumulative incidences computed from the joint model, we considered an ApoE- man who graduated from primary school and entered the cohort at 68 years old. He died at 89 years. The predicted cumulative incidence of dementia and death without dementia are plotted from the landmark age of 85 years old in figure \ref{predindiv}. From cognitive measures collected until 85 years old, this man had a probability of dying before 90 years old and before experiencing a dementia of 36.0\% [25.9\%,46.3\%]. He also had a probability of experiencing a dementia first of 16.4\% [9.1\%,29.4\%].

\section{Simulation study}

We performed a simulation study to validate the estimation program \texttt{HETMIXSURV}. We mimicked the application with two markers $Y_1$ in $[0,40]$ for which a 5-knot splines transformation was assumed and $Y_2$ in $[0,10]$ for which a threshold model was considered. We assumed 2 equiprobable latent classes with a class-specific linear latent process decline according to 65-year-centered \texttt{age}:
\begin{equation}
\begin{split}
&\Lambda(\texttt{age})|_{c_i=g} = u_{0i(g)}  + u_{1i(g)} \times \texttt{age}  \text{   with   } u_{i(g)}=(u_{0i(g)},u_{1i(g)})^\top \sim \mathcal{N}(\mu_g,B)\\
&Y_{1ij}=H_1(~\Lambda(\texttt{age}_{1ij}) +  \epsilon_{1ij}~,~\eta_{1}~)\\
&Y_{2ij}=H_2(~\Lambda(\texttt{age}_{2ij}) + \epsilon_{2ij}~,~\eta_{2}~) \text{   with   } \epsilon_{lij} \sim \mathcal{N}(0,\sigma_{\epsilon_l}^2), ~ l=1,2
\end{split}
\end{equation}
and a cause-and-class-specific Weibull model adjusted for $X_1$ with the hazard for cause $p$ in class $g$
\begin{equation}
\alpha_p(t)|_{c_i=g} = \alpha_{0pg}(t;\nu_{pg})e^{\zeta_{p} X_{1i}}
\end{equation}

Data for 500 subjects were generated as follows: the time of entry in the cohort $T_{0i}$ was generated according to a uniform distribution between 65 and 85 years old. The latent class $c$ and $X_1$ were generated from Bernoulli distributions with probability 0.5 and 0.6 respectively. Inside each latent class, the survival models were used to generate times of events $T^*_{ip}$ for each cause $p$ in $\{1,2\}$ until all the times of event were greater than $T_{0i}$. By assuming a follow-up of 20 years, the censoring time $\tilde{T}_i=T_{0i}+20$ and the observed time of event $T_{i}=\min(\tilde{T}_i,T^*_{i1},T^*_{i2})$.
Visit times were generated every 2.5 years from  $T_{0i}$ until $T_{i}$. $Y_1$ and $Y_2$ repeated measures were generated according to the model described above. The generating parameters were chosen from the application.

\begin{table}[h]
\begin{center}
\begin{tabular}{ccccccc}
\hline
model & parameter & true & mean & empirical & mean asymptotic & coverage \\
& name & value & estimate & standard deviation  & standard error & rate (95\%) \\
\hline
Class membership (logit)	&	$\xi_1$	&	0.000	&	-0.061	&	0.195	&	0.194	&	92.3	\\
Weibull for cause 1 (log)	&	$\nu_{111}$	&	4.610	&	4.630	&	0.133	&	0.122	&	87.2	\\
	&	$\nu_{211}$	&	2.750	&	2.706	&	0.637	&	0.536	&	93.5	\\
	&	$\nu_{112}$	&	4.580	&	4.580	&	0.017	&	0.015	&	92.9	\\
	&	$\nu_{212}$	&	2.730	&	2.731	&	0.185	&	0.173	&	94.7	\\
Weibull for cause 2 (log)	&	$\nu_{121}$	&	4.460	&	4.461	&	0.005	&	0.005	&	94.5	\\
	&	$\nu_{221}$	&	2.990	&	3.012	&	0.091	&	0.084	&	92.1	\\
	&	$\nu_{122}$	&	4.520	&	4.519	&	0.013	&	0.012	&	94.9	\\
	&	$\nu_{222}$	&	2.310	&	2.306	&	0.121	&	0.118	&	95.5	\\
$X_1$ on event	&	$\zeta_{1}$	&	-0.280	&	-0.314	&	0.289	&	0.279	&	93.1	\\
	&	$\zeta_{2}$	&	0.650	&	0.660	&	0.117	&	0.112	&	94.7	\\
Latent process (intercept)	&	$\mu_{01}$	&	0.000	&	-	&	-	&	-	&	-	\\
	&	$\mu_{02}$	&	-0.570	&	-0.577	&	0.202	&	0.202	&	93.5	\\
Latent process (slope)	&	$\mu_{11}$	&	-0.420	&	-0.425	&	0.075	&	0.074	&	93.9	\\
	&	$\mu_{12}$	&	-1.300	&	-1.315	&	0.101	&	0.100	&	94.3 	\\
I-splines ($Y_1$)	&	$\eta_{11}$	&	-7.030	&	-7.294	&	0.573	&	0.585	&	94.3	\\
	&	$\eta_{21}$	&	1.270	&	1.383	&	0.103	&	0.121	&	90.5	\\
	&	$\eta_{31}$	&	1.360	&	1.310	&	0.089	&	0.093	&	92.3	\\
	&	$\eta_{41}$	&	1.580	&	1.603	&	0.064	&	0.064	&	93.9	\\
	&	$\eta_{51}$	&	1.130	&	1.139	&	0.051	&	0.053	&	95.1	\\
	&	$\eta_{61}$	&	0.920	&	0.909	&	0.077	&	0.083	&	97.2	\\
	&	$\eta_{71}$	&	1.460	&	1.494	&	0.078	&	0.090	&	96.6	\\
Thresholds ($Y_2$)	&	$\eta_{12}$	&	-4.520	&	-4.572	&	0.376	&	0.364	&	92.7	\\
	&	$\eta_{22}$	&	0.680	&	0.687	&	0.041	&	0.040	&	94.3	\\
	&	$\eta_{32}$	&	0.750	&	0.753	&	0.038	&	0.038	&	95.7	\\
	&	$\eta_{42}$	&	0.640	&	0.643	&	0.034	&	0.034	&	93.7	\\
	&	$\eta_{52}$	&	0.620	&	0.622	&	0.032	&	0.032	&	95.5	\\
	&	$\eta_{62}$	&	0.600	&	0.601	&	0.031	&	0.031	&	95.1	\\
	&	$\eta_{72}$	&	0.700	&	0.704	&	0.031	&	0.032	&	95.5	\\
	&	$\eta_{82}$	&	0.640	&	0.642	&	0.030	&	0.031	&	95.1	\\
	&	$\eta_{92}$	&	0.810	&	0.815	&	0.035	&	0.035	&	94.3	\\
	&	$\eta_{102}$	&	0.810	&	0.812	&	0.035	&	0.035	&	95.9	\\
B (Cholesky)	&	$B_{00}$	&	1.000	&	-	&	-	&	-	&	-	\\
	&	$B_{01}$	&	-0.090	&	-0.085	&	0.048	&	0.046	&	89.6	\\
	&	$B_{11}$	&	0.210	&	0.191	&	0.082	&	0.060	&	86.2	\\
Residual standard-errors	&	$\sigma_{\epsilon_1}$	&	0.860	&	0.873	&	0.064	&	0.063	&	93.5	\\
	&	$\sigma_{\epsilon_2}$	&	1.270	&	1.283	&	0.095	&	0.095	&	93.5	\\

\hline
\end{tabular}
\end{center}
\caption{Simulation results for 492 converged models and samples of 500 subjects. 
Are presented, the generating parameter value, the mean estimate, the empirical standard deviation, the mean of the asymptotic standard error and the 95\% coverage rate.}\label{simu}
\end{table}

The estimation was performed for 500 replicates among which 492 converged in fewer than 25 iterations with the three criteria below 0.001 and a Gaussian quadrature with 15 points. The mean number of repeated measures per outcome was 5.4, a mean of 12.3 and 43.5 events of cause 1 were observed in each class and a mean of 226.0 and 170.9 events of cause 2 were observed. Both outcome distributions were as asymmetric as observed in the application sample. Table \ref{simu} provides the results of the simulations. The parameters were well estimated with coverage rates close to the 95\% nominal value except for the two Cholesky parameters of the random-effect covariance matrix and one Weibull parameter for cause 1 (in the class including only 12.3 events of cause 1 in mean), for which the coverage rate was slightly underestimated. \change{Note that additional simulations shown in supplementary materials also confirmed the correct estimation of other parameters such as contrasts $\gamma_k$ ($k=1,...K-1$) and standard errors of marker-specific random intercepts $\sigma_{\upsilon_k}$ ($k=1,...K$).} In this simulation setting, we also computed the type-I error of the score test statistic. Among the 492 converged models, score tests globally rejected the conditional independence assumption at the nominal value of 5\% with a proportion of 5.7\% for the global test, 9.1\% for cause 1 and 5.7\% for cause 2, specifically indicating a slightly conservative test for cause 1 that could be explained by the small number of events. With samples of 1000 subjects, proportions of rejections were 4.4\%, 4.0\% and 5.6\%.

\section{Concluding remarks}

We propose a joint model for multiple longitudinal outcomes and multiple times-to-event. A few studies aimed at modeling multivariate longitudinal outcomes and multiple events \cite{chi2006,li2012,andrinopoulou2014}, but none of them relied on a latent class approach. They all used the shared random-effect approach which may be of particular interest for assessing specific assumptions regarding the dependency between the repeated longitudinal outcomes and the times-to-event. However, as shown in the more standard univariate context \cite{proust-lima2014}, the joint latent class approach may provide a much better fit to the data. Indeed, on the longitudinal side, it models several mean profiles of trajectories whereas the shared random-effect models estimate a unique mean trajectory. On the survival side, it may be explained by the fact that the joint latent class model estimates a stratified risk function instead of a unique risk function with proportional continuous association. 
\change{The main drawback of the joint latent class approach is that models must be repeatedly estimated with different numbers of classes and different initial values because the likelihood of mixture models may be multi-modal. This is what we did in the application by estimating each model from different random initial values and going back and forth from models with different $G$. In our experience, multi-modality, which corresponds to suboptimal maxima, often arises when the optimum number of classes is not reached.}

By relying on a unique latent process that generated the set of multivariate measures, the model may seem very specific. However, it applies to various settings. In particular, with the recent interest in patient-reported outcomes and more generally indirect measures of psychological or biological processes (quality of life, well-being, immunological response) in chronic diseases, it provides a complete approach to describe developmental trajectories in association with clinical events or dropout. It includes, for example, the 2-parameter IRT longitudinal model dedicated to the analysis of questionnaires data as a special case \cite{proust-lima2012}.

\change{The model was developed in a parametric but flexible framework. In the repeated longitudinal markers model for example, any function of time, including splines or fractional polynomials can be considered to approximate the trajectory of the latent process over time. Along with the Beta cumulative distribution functions that are already relatively flexible, approximations by splines and the non-parametric threshold were also considered for the link functions between the latent process and the longitudinal markers. In the survival part, we also proposed different types of baseline risk functions including M-splines and piecewise exponential functions.}

Such a joint model is of particular interest to provide dynamic individual predictions in competing settings. Although we showed how to compute individual predictions, we did not evaluate its predictive performances. Methods for evaluating the predictive ability (Brier score and ROC curve) of joint models in a competing setting can be found elsewhere \cite{blanche2014}.

Although dementia and death constitute competing events, multiple \change{non-competing} events could be of interest in other contexts (like different types of recurrences in cancer). By considering $P$ observed times $T_{ip}=\min(\tilde{T}_i,T^*_{ip})$ and $P$ indicators $\delta_{ip}=1$ if the clinical event of nature $p$ occurred before censoring instead of the unique observed time $T_i$ and indicator $\delta_{i}$, and by replacing equation \eqref{liksurv} in the likelihood formula by $f(T_{ip},\delta_{ip} \mid c_i = g;\theta_G)=e^{-\sum_{p=1}^P A_p(T_{ip} \mid c_i=g;\theta_G)}\prod_{p=1}^P \alpha_p(T_{ip} \mid c_i=g;\theta_G)^{1_{\delta_{ip}=1}}$, this approach can also handle multiple correlated times of events instead of competing events.

\section{Appendix: conditional density of ordinal and bounded outcomes}

For outcomes $k$ with values in $[0,M_k]$ that are modelled using a threshold transformation, the conditional density of $Y_{kij}$ for subject $i$ and occasion $j$ used in equation \eqref{condY} is:
\begin{equation}
\hspace*{-1.5cm} f_y(Y_{kij} | c_i = g,u_{i(g)},\upsilon_{ki}) = \prod_{m=0}^{M_k} \left( \phi \left (\frac{\eta_{m+1}-E(\tilde{Y}_{kij}|u_{i(g)},\upsilon_{ki},c_i = g)}{\sigma_{\epsilon_k}} \right)-\phi \left (\frac{\eta_{m}-E(\tilde{Y}_{kij}|u_{i(g)},\upsilon_{ki},c_i = g)}{\sigma_{\epsilon_k}} \right ) \right )^{I_{Y_{kij}=m}}
\end{equation}
with $E(\tilde{Y}_{kij}|u_{i(g)},\upsilon_{ki},c_i = g)= X_{\Lambda i}(t_{kij}) \beta_g +Z_{i}(t_{kij}) u_{i(g)} + X_{Yi}(t_{kij}) \gamma_k+\upsilon_{ki}$, $\eta_{M_k+1}=+\infty$ and $\eta_{0}=-\infty$.

For bounded outcomes $k$ in $[0,M_k]$, the conditional density of $Y_{kij}$ used in equation \ref{condY} is:
\begin{equation}
\begin{split}
f_y(Y_{kij} | c_i = g,u_{i},\upsilon_{ki}) &=  \left ( \phi \left (\frac{H_k^{-1}(0)-E(\tilde{Y}_{kij}|u_{i(g)},\upsilon_{ki},c_i = g)}{\sigma_{\epsilon_k}} \right)\right) ^{I_{Y_{kij}=0}} \times \left( 1 - \phi \left (\frac{H_k^{-1}(M_k)-E(\tilde{Y}_{kij}|u_{i(g)},\upsilon_{ki},c_i = g)}{\sigma_{\epsilon_k}} \right) \right)^{I_{Y_{kij}=M_k}}\\
&~~~ \times \left(  \frac{1}{\sqrt{2\pi}\sigma_{\epsilon_k}} \text{exp} \left (- \frac{\left (\tilde{Y}_{kij}-E(\tilde{Y}_{kij}|u_{i(g)},\upsilon_{ki},c_i = g) \right )^2 }{2 \sigma_{\epsilon_k}^2} \right ) J(H_k^{-1}(Y_{kij})) \right)^{I_{Y_{kij}\in (0,M_k)}}
\end{split}
\end{equation}
{The univariate integral over $\upsilon_{ki}$ (for $k=1,...,K_1$) and the multivariate integral over $u_i$ are evaluated numerically using Gaussian quadratures.}



\section*{Acknowledgments}
Computer time was provided by the computing facilities MCIA (M\'esocentre de Calcul Intensif Aquitain) at the Universit\'e de Bordeaux and the Universit\'e de Pau et des Pays de l'Adour. The PAQUID study was funded by SCOR Insurance, Agrica, the Conseil R\'egional of Aquitaine, the Conseils G\'en\'eraux of Gironde and Dordogne, the Caisse Nationale de Solidarit\'e pour l'Autonomie (CNSA), IPSEN, the Mutualit\'e Sociale Agricole (MSA), and Novartis Pharma (France).
{\it Conflict of Interest}: None declared.

\bibliographystyle{biom}
\bibliography{Ma_bibliotheque}

\end{document}